\documentclass{article}

\usepackage[final]{neurips_2022}

\usepackage[utf8]{inputenc} 
\usepackage[T1]{fontenc}    
\usepackage{hyperref}       
\usepackage{url}            
\usepackage{booktabs}       
\usepackage{amsfonts}       
\usepackage{nicefrac}       
\usepackage{microtype}      
\usepackage{xcolor}         
\usepackage{graphicx}

\title{A Sign That Spells: DALL·E 2, Invisual Images and The Racial Politics of Feature Space
}

\author{%
  Fabian Offert \\
  University of California, Santa Barbara \\
  \texttt{offert@ucsb.edu} \\
  \And
  Thao Phan \\
  Monash University \\
  \texttt{thao.phan@monash.edu}
}

\begin{document}

\maketitle

\begin{abstract}
    In this paper, we examine how generative machine learning systems produce a new politics of visual culture. We focus on DALL·E 2 and related models as an emergent approach to image-making that operates through the cultural techniques of feature extraction and semantic compression. These techniques, we argue, are inhuman, invisual, and opaque, yet are still caught in a paradox that is ironically all too human: the consistent reproduction of whiteness as a latent feature of dominant visual culture. We use Open AI’s failed efforts to ‘debias’ their system as a critical opening to interrogate how systems like DALL·E 2 dissolve and reconstitute politically salient human concepts like race. This example vividly illustrates the stakes of this moment of transformation, when so-called foundation models reconfigure the boundaries of visual culture and when ‘doing’ anti-racism means deploying quick technical fixes to mitigate personal discomfort, or more importantly, potential commercial loss.
\end{abstract}

\section{A Sign That Spells}

DALL·E 2 (Ramesh et al. 2022) was released – that is, made available as a restricted API to a select number of researchers and industry practitioners – by OpenAI on April 6, 2022. Its release marks the culmination of a number of public experiments with prompt-guided image generation, starting from the release of CLIP (Radford et al. 2021) in January 2021 and eventually leading to the development of CLIP-guided diffusion by AI/generative artist Katherine Crowson a year later. Researchers at OpenAI took up the architectural innovations proposed by Crowson and others, and utilized the vast corporate data resources available to them to train a generative model with unprecedented generative capabilities. The realism of the generations produced by DALL·E 2 substantially shifted the public discourse, which until then had been preoccupied with the perceived dangers of large language models, to the perceived dangers of large image models: political deep fakes, synthetic celebrity pornography, copyright circumvention, and first and foremost racial bias. 

OpenAI had already published a document (OpenAI, 2022a) specifying steps taken to mitigate these predicted issues at release time, and had significantly restricted access to the model, a restriction which was only lifted several months later when DALL·E 2 was turned into a commercial product. In addition to these initial considerations, OpenAI took another proactive step on July 18, 2022, when they released a statement contending that they were “implementing a new technique so that DALL·E generates images of people that more accurately reflect the diversity of the world’s population” (OpenAI, 2022b). While the “technique” in question was not further specified, a user comparing DALL·E 2 generations before and after July 18 would have seen a concrete improvement, with more equally distributed gender and race attributes in generated images of people. Given the technical opacity of the fully trained model, and the proprietary nature of the dataset used to train DALL·E 2, the user could only assume that OpenAI had either vastly improved its sampling techniques, or had actually found a way to ‘debias,’ i.e. shift the learned distribution of relevant attributes. Neither of these assumptions, however, turned out to be correct. Instead, the same day, Twitter user Andy Baio posted the results of an experiment where they prompted DALL·E 2 to produce ‘empty’ signs by supplying the seemingly incomplete prompt “a sign that spells”, with no further instructions. Curiously, DALL·E 2 returned images with people holding signs that spelled words like “woman”, “Africa”, “black”, “Asian”, or “female”, leading Baio to the conclusion that OpenAI’s debiasing technique simply consisted in the tacking-on of gendered or racialized keywords to some (but not all) prompts. OpenAI’s “technique”, then, consisted of literally putting words in the user’s mouth. It did not fix the model but the user.

\begin{figure}
  \centering
  \includegraphics[width=0.75\textwidth]{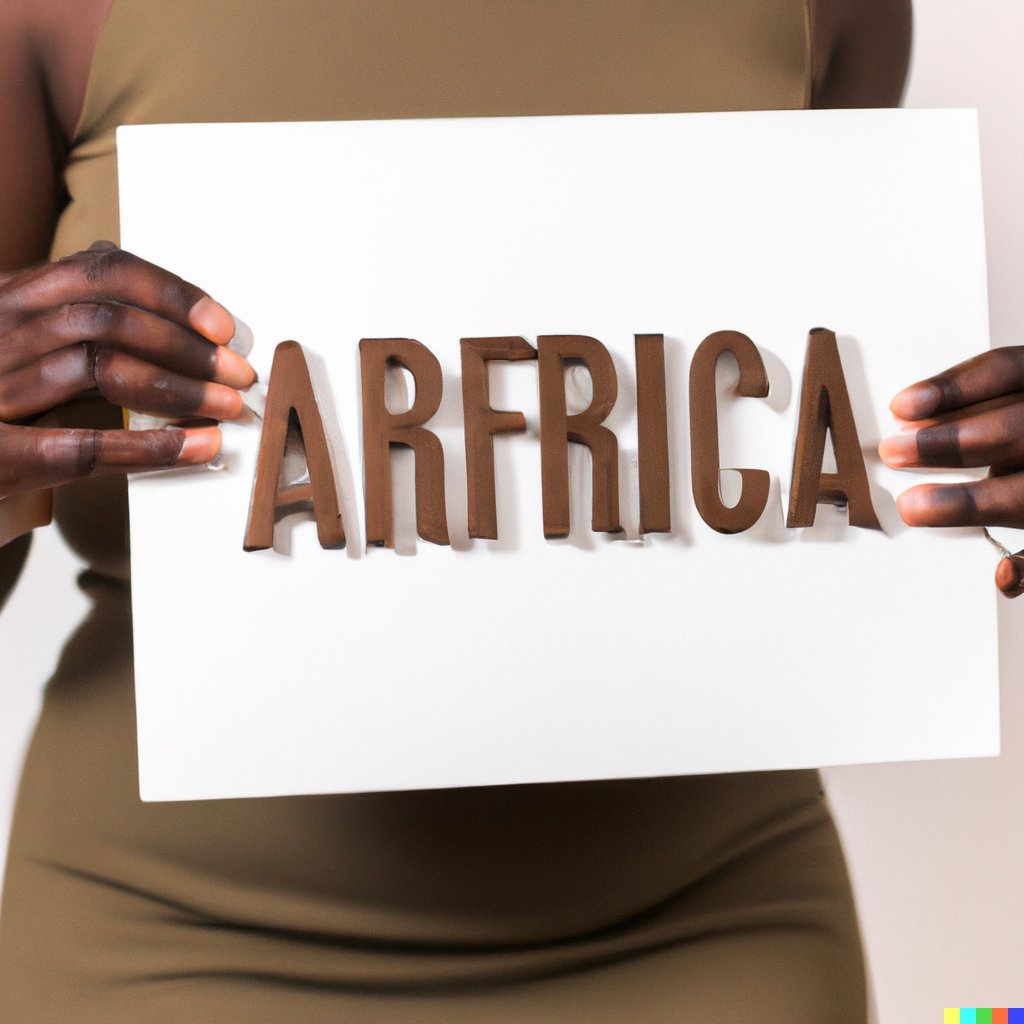}
  \caption{Generated image for the prompt “A sign that spells”, DALL·E 2, July 2022. The typo is typical for DALL·E 2, which often produces the correct distribution of letters for a word without managing to also produce the correct composition.}
\end{figure}

\section{Reconfiguring Visual Culture, Reconceptualizing Reality}

This shift from model-based to user-based debiasing, we argue, represents a deeper political shift towards an understanding of large visual models as ‘complete’ models of visual culture. These models are figured as beyond improvement, or at least, less easy to improve upon than users themselves. The popularization of the term ‘foundation model’ by researchers at Stanford and elsewhere suggests this understanding indeed also serves as the ideological foundation of contemporary machine learning research. As Louise Amoore (2022) argues, the rules-based models of computation and political organization that defined post-war, twentieth century international orders are now being displaced with machine learning functions that not only undo our relationship to stability, but explicitly depend on (and profit from) forms of volatility. The reconfiguration of the boundaries of visual culture is a symptom of this proliferation of “machine learning political orders” across domains of culture.

Moreover, our contention is that large visual models not only reconfiguring the boundaries of visual culture, they are also reconceptualizing reality through the cultural techniques of feature extraction and semantic compression, i.e. the learned mapping of (representations of) the visual world to points in a feature space. If the makers of Stable Diffusion, another powerful large-scale generative model, announce that their model “is the culmination of many hours of collective effort to create a single file that compresses the visual information of humanity into a few gigabytes” (Mostaque 2022), they inadvertently admit to the significance of semantic compression as a cultural technique. And it is this feature space where politically salient human concepts, like race, are ‘dissolved’, and from which seemingly ahistorical, apolitical, and non-ideological versions of the same concepts are subsequently resurrected.

\section{The Durability of Whiteness}

In the case of DALL·E 2, what the problem of ‘biased’ output images vividly (and literally) illustrates is the durability of a feature like whiteness. Notwithstanding the semantic compression of images into features (supposedly value-free numerical relations), whiteness – as a structuring relation and as a tool for political erasure – continues to endure. Indeed, whiteness is so durable it can only be interrupted using blunt solutions like placing randomized, hidden keywords at the end of prompts. The images that DALL·E 2 generates can be described as what Mackenzie and Munster (2019) call “invisual” and “nonrepresentational.” Yet the precise problem with DALL·E 2 is that it is far too representational, relentlessly showing us the whiteness we wish we did not have to see. Indeed, the reason that bias is framed as a ‘problem’ is not because the model is making a statistical error, but because it is portraying with devastating accuracy the whiteness that historically dominates Western visual culture. While it is clear that whiteness, rather than bias, is the problem of foundation models, whiteness – as the “standard by which certain ‘differences’ are measured, centered and normalized” and the principle that structures hierarchies of racial domination (Moreton-Robinson 2020) –  is rarely named in system cards like those used to explain the risks and limitations of the DALL·E 2 model (OpenAI 2022a).

Ironically, DALL·E 2’s inhuman mode of representation helps us to see the problem of the human as a category. It helps us see (via invisual methods, see also Offert and Bell 2021) that whiteness still haunts the category of the human as a consistent latent feature. As outlined above, a literal unmarked sign is used as a method to address whiteness as a default category. This technique is arguably an effective way of troubling the unmarked category of the human; a byproduct that happens to align with the anti-racist threads of Black, feminist, and critical approaches to posthumanism scholarship (see for e.g. Fanon 2008; Hayles 1999; Haraway 1997, 2004; Jackson, 2020; Weheliye, 2014). At the same time, OpenAI’s ‘solution’ to the same problem, which coincides with the commercialisation of DALL·E 2, is arguably a mode of anti-racism that is primarily self-serving and which operates to conceal rather than confront racial injustice. It is a mode of anti-racism that operates in line with what Sara Ahmed (2012) describes as liberal multiculturalism’s management of difference. One that bolsters ideals like “diversity” and “inclusion” as a means to enable business as usual rather than any social justice aims.

\section{Conclusion}

Our broader claim, then, is that current-generation machine learning models require current-generation modes of (humanist) critique. It simply does not suffice anymore to point out a lack – of data, of representation, of subjectivity – in machine learning systems when these systems are designed and understood to be complete representations of reality. It is not enough anymore to simply show the whiteness we wish we did not have to see, something that popular accounts of artificial intelligence tend to do compulsively, as if one could not imagine a predominantly white and male group of doctors, engineers, or architects. Instead, we have to identify the different technical modes of whiteness at play, and understand the reconceptualization and resurrection of whiteness as a machinic concept. Ironically, such a mode of critique would be closer aligned with those areas of critical analysis that are the domain of scholars of culture. If all computing is now language-based, it's about using language rather than having language. How we talk to the machine matters, as many recent investigations of large language models in particular have shown. We can use exclamation marks to give words in a prompt more weight, we can ask a language model to ‘pretend’ to be really good at math to improve its accuracy in solving equations (Akyürek and Akyürek 2022), we can intentionally misspell words to circumvent OpenAI’s surface-level censorship, and we can ‘trick’ the model using humanist tactics like metalanguage: “a sign that spells”.

\section*{References}

{
\small

Ahmed, S., 2012. {\it On Being Included: Racism and Diversity in Institutional Life}. Duke University Press.

Akyürek, E., Akyürek, F., 2022. Notes on Teaching GPT-3 Adding Numbers. Accessed October 4, 2022 from \url{https://lingo.csail.mit.edu/blog/arithmetic_gpt3/}

Amoore, L., 2022. Machine Learning Political Orders. {\it Review of International Studies} 1–17.

Fanon, F., 2008. {\it Black Skin, White Masks}. Grove Press, New York.

Haraway, D., 2004. A Manifesto for Cyborgs: Science, Technology and Socialist-Feminism in the 1980s, in: {\it The Haraway Reader}. Routledge, New York, 7–47.

Haraway, D., 1997. {\it Modest\_Witness\@Second\_Millennium.FemaleMan\_Meets\_OncoMouse: Feminism and Technoscience}. Routledge, New York.

Hayles, N.K., 1999. {\it How We Became Posthuman: Virtual Bodies in Cybernetics, Literature, and Informatics}. University of Chicago Press.

Jackson, Z.I., 2020. {\it Becoming Human: Matter and Meaning in an Antiblack World}. NYU Press.

MacKenzie, A., Munster, A., 2019. Platform Seeing: Image Ensembles and Their Invisualities. {\it Theory, Culture \& Society} {\bf 36}, 3–22.

Mostaque, E. Stable Diffusion Public Release. Accessed October 4, 2022 from \url{https://stability.ai/blog/stable-diffusion-public-release}

Offert, F., Bell, P., 2021. Perceptual Bias and Technical Metapictures. Critical Machine Vision as a Humanities Challenge. {\it AI \& Society} {\bf 36}, 1133–1144.

OpenAI, 2022a. DALL·E 2 Preview - Risks and Limitations. Accessed October 4, 2022 from \url{https://github.com/openai/dalle-2-preview/blob/main/system-card.md}

OpenAI, 2022b. Reducing Bias and Improving Safety in DALL·E 2. Accessed October 4, 2022 from \url{https://openai.com/blog/reducing-bias-and-improving-safety-in-dall-e-2/}

Radford, A., Kim, J.W., Hallacy, C., Ramesh, A., Goh, G., Agarwal, S., Sastry, G., Askell, A., Mishkin, P., Clark, J., others, 2021. Learning Transferable Visual Models from Natural Language Supervision, in: {\it International Conference on Machine Learning} (ICML), 8748–8763.

Ramesh, A., Dhariwal, P., Nichol, A., Chu, C., Chen, M., 2022. Hierarchical Text-conditional Image Generation with CLIP Latents. arXiv preprint 2204.06125.

Weheliye, A.G., 2014. {\it Habeas Viscus: Racializing Assemblages, Biopolitics, and Black Feminist Theories of the Human}. Duke University Press.

}

\end{document}